\documentclass[pdflatex,sn-mathphys-num]{sn-jnl}

\usepackage{graphicx}%
\usepackage{multirow}%
\usepackage{amsmath,amssymb,amsfonts}%
\usepackage{amsthm}%
\usepackage{mathrsfs}%
\usepackage[title]{appendix}%
\usepackage{xcolor}%
\usepackage{textcomp}%
\usepackage{manyfoot}%
\usepackage{booktabs}%
\usepackage{algorithm}%
\usepackage{algorithmicx}%
\usepackage{algpseudocode}%
\usepackage{listings}%

\usepackage{float}

\usepackage{tabularx}
\usepackage{pifont}
\usepackage[table]{xcolor}

\theoremstyle{thmstyleone}%
\theoremstyle{thmstyletwo}%

\theoremstyle{thmstylethree}%

\begin{document}


\title[Article Title]{Meta-optical processors for broadband complex-field image operations}

\author*[1]{\fnm{Linzhi} \sur{Yu}}\email{linzhi.yu@tuni.fi}

\author[1]{\fnm{Jesse} \sur{Pietila}}\email{jesse.pietila@tuni.fi}

\author[1]{\fnm{Haobijam J.} \sur{Singh}}\email{johnson.singh@tuni.fi}

\author[1]{\fnm{Arttu} \sur{Nieminen}}\email{arttu.nieminen@tuni.fi}

\author*[1,2]{\fnm{Humeyra} \sur{Caglayan}}\email{h.caglayan@tue.nl}

\affil*[1]{\orgdiv{Department of Physics}, \orgname{Tampere University}, \city{Tampere}, \postcode{33720}, \country{Finland}}

\affil[2]{\orgdiv{Department of Electrical Engineering, Photonic Integration Group}, \orgname{Eindhoven University of Technology}, \city{Eindhoven}, \postcode{5600 MB}, \country{The Netherlands}}

\abstract{All-optical image processing provides a fast and energy-efficient alternative to conventional electronic systems by directly manipulating optical wavefronts. However, metasurface-based optical processors reported to date are often limited in functionality, operating bandwidth, or input modality, which restricts their adaptability across different image processing tasks. Here, we demonstrate a broadband metasurface platform capable of performing diverse analog image processing operations on both amplitude- and phase-encoded inputs. This platform is realized using a single-layer dielectric metasurface designed through an end-to-end, task-driven inverse design framework. By tailoring the spatial-frequency components of incident image wavefronts, the metasurface implements analog operations such as edge detection and pattern recognition across a 200~nm wavelength bandwidth in the visible spectrum. Furthermore, we develop a compact processor architecture that integrates imaging and computation within a reduced optical footprint. These results establish a flexible and compact metasurface-based optical processor with strong potential for integration into practical imaging and optical computing systems.}

\maketitle

\section*{Introduction}\label{sec1}

High-speed and energy-efficient image processing is increasingly critical for applications ranging from machine vision and autonomous systems to scientific imaging and instrumentation. Conventional digital image processing pipelines rely on analog-to-digital conversion followed by sequential electronic computation, which introduces latency, incurs substantial power consumption, and faces growing scalability challenges as transistor technologies approach their physical limits~\cite{williams2017s}. During optical-to-electronic conversion, only the intensity of light is typically recorded, while phase information encoded in the optical field is discarded, resulting in a loss of information richness. These limitations have motivated the exploration of optical analog image processing, in which spatial information is processed directly in the optical domain. By avoiding digital conversion and exploiting the intrinsic parallelism of light propagation, optical approaches offer compelling advantages in processing speed, energy efficiency, and information capacity~\cite{solli2015analog,mcmahon2023physics}.

Conventional optical image processors typically rely on cascaded bulky refractive or reflective components, which hinder miniaturization, scalability, and system-level integration~\cite{javidi1994real,chang2018hybrid,wang2023image,kalinin2025analog}. Recent advances in metasurfaces, planar optical elements composed of arrays of engineered scatterers, have enabled unprecedented control over optical wavefronts with subwavelength spatial resolution in an ultrathin form factor, making them well suited for compact analog image processing~\cite{yu2011light,sun2012gradient,lin2014dielectric,kuznetsov2024roadmap,brongersma2025second}.

Building on these capabilities, metasurface-based analog image processing has been demonstrated for operations such as edge detection~\cite{zhou2019optical,zhou2021two,wang2023single,yu2025multifunctional,qiu2025metasurface,deng2024broadband,huo2020photonic,kim2022spiral,huo2024broadband,guo2018photonic,kwon2018nonlocal,zhou2020flat,cotrufo2024reconfigurable,yao2025nonlocal,pearson2025inverse,wang2022single,fu2022ultracompact,yu2025meta,tanriover2023metasurface,swartz2024broadband}, image denoising~\cite{icsil2024all,fu2022ultracompact,pearson2025inverse}, and acceleration of artificial neural networks~\cite{lin2018all,li2021spectrally,luo2022metasurface,liu2022programmable,chen2023all,zheng2022meta,zheng2024multichannel,fu2024optical}. However, most reported implementations focus on realizing a specific functionality, typically edge detection, which limits their generalizability to broader image processing tasks. In addition, many demonstrated processors operate only at a single wavelength, constraining their applicability across the visible spectrum, which underpins imaging technologies including cameras, displays, and microscopy. Most existing approaches also address either amplitude- or phase-encoded inputs only, rather than enabling a common processing functionality to operate on both modalities within the same metasurface architecture, further limiting overall versatility.

In this work, we present a metasurface-based platform, termed meta-optical processors, capable of broadband all-optical image processing for diverse tasks. These processors are implemented using a single-layer dielectric metasurface architecture designed through an end-to-end, task-driven inverse design framework. By tailoring the spatial-frequency components of incident image wavefronts, this approach enables a common processing functionality to operate on both amplitude- and phase-encoded inputs. Image edge detection and letter pattern recognition are experimentally demonstrated over a wavelength range spanning 430~nm to 630~nm. To reduce system complexity, a compact processor architecture is introduced that integrates imaging and computation within a single planar layer, eliminating the need for conventional multi-element $4f$ systems. These results establish a scalable and versatile platform for optical analog image processing, combining functional flexibility, broadband operation, and compact integration.

\section*{Results}\label{sec2}

\subsection*{Principle and design of the meta-optical processor}\label{subsec2_1}

\begin{figure}[H]
\centering
\includegraphics[width=0.85\textwidth]{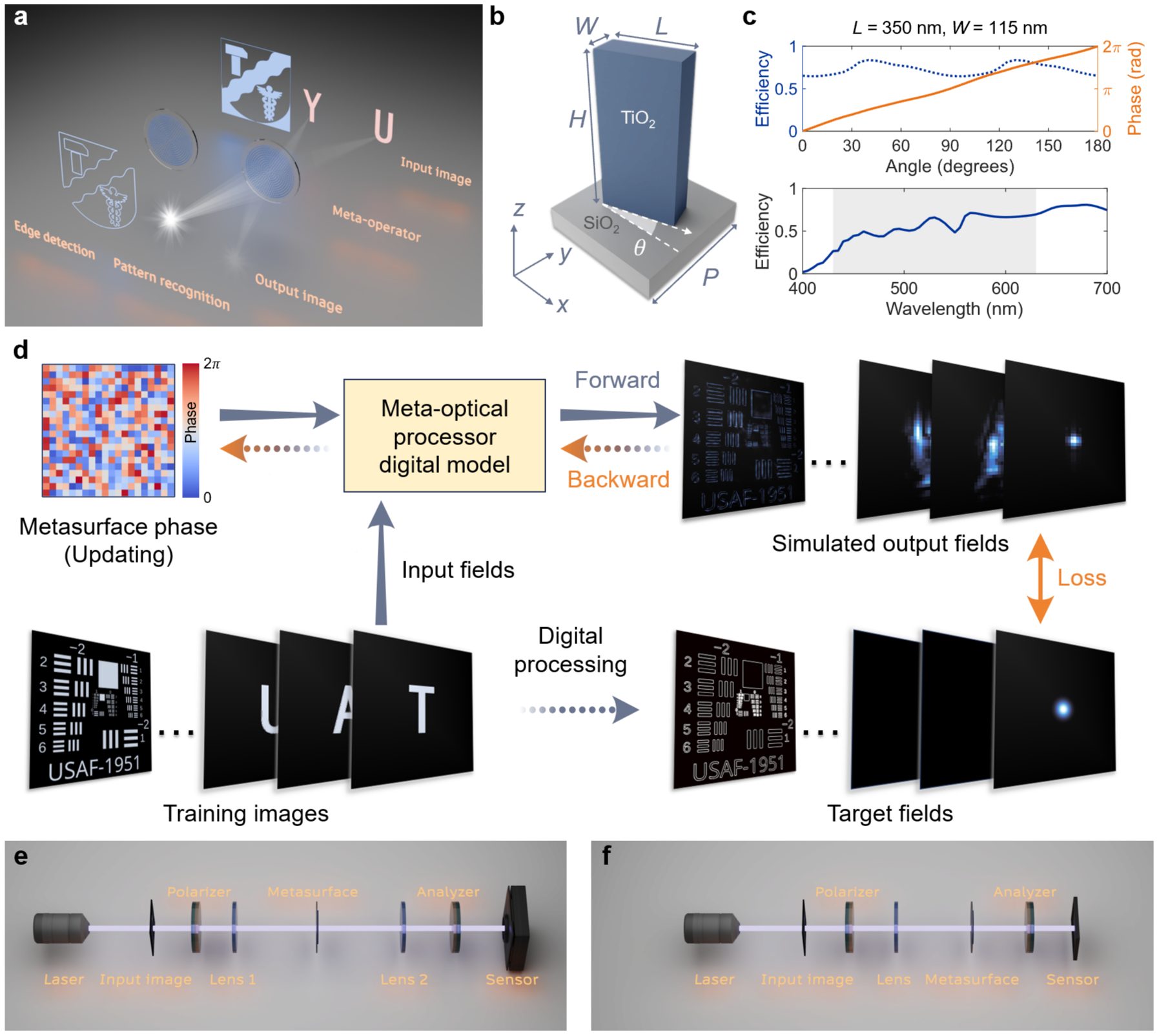}
\caption{\textbf{Concept and design of meta-optical processors for complex-field image processing.} 
\textbf{a} Schematic of a meta-optical processor performing all-optical complex-field image edge detection and letter pattern recognition. 
\textbf{b} Unit-cell design of birefringent TiO$_2$ nanopillar meta-atoms on a SiO$_2$ substrate, defined by a height $H = 600\,\mathrm{nm}$, length $L = 350\,\mathrm{nm}$, width $W = 115\,\mathrm{nm}$, period $P = 450\,\mathrm{nm}$, and an in-plane orientation angle $\theta$ that encodes the geometric phase. 
\textbf{c} Simulated spin-conversion efficiency and geometric phase shift as functions of the nanopillar orientation angle at 530\,nm (top), and broadband spin-conversion efficiency of the optimized meta-atoms across the 400--700\,nm spectral range (bottom). The shaded region (430--630\,nm) indicates the operational wavelength range used in this work. 
\textbf{d} Differentiable meta-optical processor model for metasurface phase optimization. A forward model simulates the modulation of incident image wavefronts by the meta-optical processor, and gradients of an output-defined loss are evaluated via automatic differentiation in backward propagation to update the phase profile.
\textbf{e} Schematics of the experimental setup for characterizing the meta-optical processors using the $4f$ imaging system. 
\textbf{f} Schematics of the experimental setup for characterizing the compact meta-optical processor.}
\label{fig 1}
\end{figure}

Meta-optical processors implement image processing by controlling the propagation of complex optical fields through a metasurface in a fully passive, all-optical manner, as illustrated in Figure~\ref{fig 1}a. Instead of relying on predefined analytical filters~\cite{zhou2019optical,zhou2021two,yu2025multifunctional,huo2020photonic,huo2024broadband,wang2022single,wang2023single,yu2025meta,zheng2024multichannel}, the metasurface phase profile is designed in a task-driven manner to globally reshape the incident image wavefront, such that the desired image transformation emerges at the output plane.

To realize this task-driven phase design, we develop an inverse design framework based on a differentiable physical model of optical wave propagation through the meta-optical processor, as illustrated in Figure~\ref{fig 1}d. In this model, the input image is represented as a complex optical field $U(x,y)$, which is transformed into the spatial-frequency domain via a two-dimensional Fourier transform,
\begin{equation}
\tilde{U}(k_x, k_y) = \mathcal{F}\left\{ U(x, y) \right\}.
\end{equation}
The metasurface is modeled as a phase-only modulator described by a spatially varying transmission function,
\begin{equation}
T(k_x, k_y) = \exp\left[i\,\phi(k_x, k_y)\right],
\end{equation}
where $\phi(k_x, k_y)$ denotes the metasurface phase profile. The modulated spectrum is given by
\begin{equation}
\tilde{U}'(k_x, k_y) = \tilde{U}(k_x, k_y)\,T(k_x, k_y),
\end{equation}
and the corresponding output optical field is obtained through an inverse Fourier transform,
\begin{equation}
U'(x, y) = \mathcal{F}^{-1}\left\{ \tilde{U}'(k_x, k_y) \right\}.
\end{equation}
The final output intensity is $I(x,y) = |U'(x,y)|^2$.

A desired image processing function is specified by a target intensity distribution $I^{\mathrm{target}}(x,y)$. The metasurface phase profile $\phi(k_x,k_y)$ is determined by minimizing a differentiable loss function $\mathcal{L}(I, I^{\mathrm{target}})$ that quantifies the discrepancy between the computed and target outputs. Gradients of the loss with respect to the phase profile are evaluated through automatic differentiation of the physical propagation model, and the optimization is performed using an iterative, gradient-based algorithm~\cite{paszke2019pytorch,kingma2014adam}. As a result, this framework enables the realization of meta-optical processors for a wide range of analog image processing functions. Additional implementation details are provided in Supplementary Information Section~1.

The metasurface is physically implemented as a periodic array of rectangular titanium dioxide (TiO$_2$) nanopillars~\cite{khorasaninejad2016metalenses} fabricated on a fused silica (silicon dioxide, SiO$_2$) substrate, as shown in Figure~\ref{fig 1}b. Each nanopillar functions as a subwavelength birefringent element with fixed geometry and a spatially varying in-plane orientation angle $\theta(x,y)$. Under circularly polarized illumination, the nanopillars act as local half-wave plates, converting the spin state of the incident light while imparting a geometric phase shift of $\pm 2\theta(x,y)$, with the sign determined by the handedness of the input polarization. This geometric-phase mechanism enables continuous $2\pi$ phase modulation across the metasurface through rotation alone, without modifying the physical dimensions of the meta-atoms~\cite{arbabi2015dielectric,balthasar2017metasurface}.

To achieve broadband operation across the visible spectrum, the nanopillar geometry was optimized using full-wave electromagnetic simulations. The selected design comprises a height $H = 600\,\mathrm{nm}$, length $L = 350\,\mathrm{nm}$, width $W = 115\,\mathrm{nm}$, and lattice period $P = 450\,\mathrm{nm}$. Figure~\ref{fig 1}c shows the simulated phase response and spin-conversion efficiency as functions of nanopillar orientation at a wavelength of 530~nm, corresponding to the center of the operational band. The broadband efficiency across 400--700~nm is also shown, with the shaded region (430--630~nm) indicating the wavelength range tested in this work. Although the conversion efficiency exhibits moderate spectral variation, the phase modulation remains intrinsically achromatic owing to the geometric-phase nature of the design. Additional simulation results and design details are provided in Supplementary Information Section~2.

\subsection*{Meta-optical complex-field edge detection}\label{subsec2_2}

\begin{figure}[H]
\centering
\includegraphics[width=0.85\textwidth]{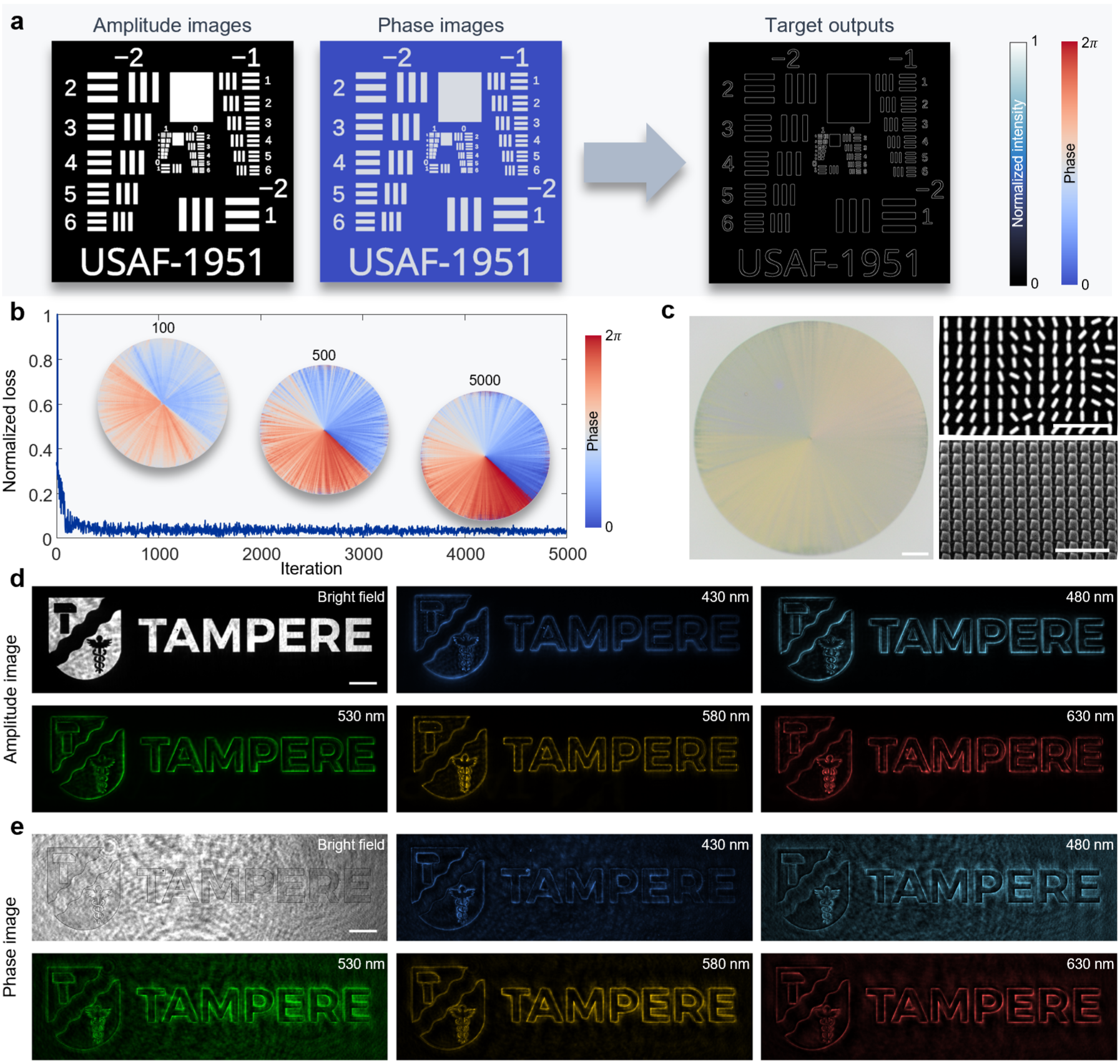}
\caption{\textbf{Experimental demonstration of meta-optical complex-field edge detection.} 
\textbf{a} Amplitude- and phase-encoded input images based on the USAF-1951 resolution target, along with the corresponding target edge-intensity distribution used for inverse design. 
\textbf{b} Optimization trajectory showing the normalized loss during 5000 iterations. Insets show the evolution of the metasurface phase profile from an initially flat distribution at iterations 100, 500, and 5000. 
\textbf{c} Optical microscope image of the fabricated metasurface and corresponding scanning electron microscopy (SEM) images of the TiO$_2$ nanopillar array, shown in top and tilted views. 
\textbf{d} Experimental edge detection results for an amplitude sample consisting of a transmission mask patterned with the Tampere city logo. 
\textbf{e} Experimental edge detection results for a phase sample formed by a transparent surface-relief structure. 
For both samples, the unprocessed outputs and the corresponding edge-detected results at wavelengths of 430, 480, 530, 580, and 630~nm are shown. Scale bars: 200~\textmu m (optical microscopy in \textbf{c}), 2~\textmu m (SEM in \textbf{c}), and 500~\textmu m (\textbf{d}, \textbf{e}).}
\label{fig 2}
\end{figure}

As a first demonstration of the proposed meta-optical framework, we implement image edge detection, a representative image processing task that provides a clear and intuitive benchmark for evaluating complex-field manipulation. The metasurface is designed to transform input images encoded in the complex optical field into edge-enhanced intensity outputs.

The metasurface phase profile is optimized using a set of reference images derived from a USAF-1951 resolution target comprising horizontal and vertical line pairs spanning a broad range of spatial frequencies. Two encoding schemes are employed to generate the input optical fields. For amplitude-encoded inputs, the field amplitude follows the normalized grayscale image with a uniform phase. For phase-encoded inputs, the field amplitude is uniform, while the phase is modulated such that bright regions correspond to a phase shift of $\pi$ and dark regions remain at zero phase. The corresponding target outputs are generated by applying a digital Canny edge detector to the original image~\cite{gonzalez2006}. To enhance robustness against feature orientation, the input-target image pairs are randomly rotated during optimization. Additional details on data preparation are provided in Supplementary Information Section~4.

The phase profile is obtained by minimizing the mean squared error between the computed and target output intensities,
\begin{equation}
\mathcal{L}_{\mathrm{edge}} = \left\| I(x, y) - I^{\mathrm{target}}(x, y) \right\|_2^2.
\end{equation}

The meta-optical processor is implemented as a circular aperture with a diameter of 4000 meta-atoms, corresponding to a physical size of 1.8\,mm. Starting from an initially uniform phase distribution, the optimization converges to a spatially varying phase profile spanning the full $[0,2\pi]$ range, which implements the desired edge-enhancing function. Figure~\ref{fig 2}b shows the optimization trajectory, showing a monotonic decrease in the normalized loss over 5000 iterations, with representative phase profiles shown at intermediate and final stages. The fabricated metasurface is shown in Figure~\ref{fig 2}c.

We evaluate the broadband edge-detection performance of the fabricated meta-optical processor using two test samples under the experimental setup shown in Figure~\ref{fig 1}e (see Methods and Supplementary Information Section~11). The first sample is a transmission mask patterned with the Tampere city logo (Supplementary Information Section~3), which serves as an amplitude-encoded image with spatially varying intensity. The second sample is a transparent surface-relief structure that introduces spatially varying phase delays while maintaining nearly uniform transmitted intensity. These two samples respectively assess the response of the processor to amplitude and phase variations in the incident optical field (see Supplementary Information Section~13).

Experimental results are shown in Figure~\ref{fig 2}d,e. For each sample, the unprocessed bright-field image is shown together with the corresponding edge-detected outputs obtained at wavelengths of 430, 480, 530, 580, and 630~nm. For the amplitude-encoded image (Figure~\ref{fig 2}d), the metasurface selectively enhances sharp intensity transitions, resulting in clear edge features that remain consistent across the entire wavelength range. For the phase-encoded image (Figure~\ref{fig 2}e), the unprocessed image exhibits minimal contrast, as expected for a phase object under direct illumination. After metasurface processing, the edge-enhanced outputs clearly reveal the contours of the surface-relief pattern at all wavelengths, demonstrating the processor’s ability to extract spatial features that are otherwise inaccessible in standard intensity imaging. Although the edge contrast at 430~nm is slightly reduced due to lower spin-conversion efficiency and residual background from the experimental setup, the edge features remain clearly resolved. Additional simulation results and comparative analyses are provided in Supplementary Information Section~5.

\subsection*{Meta-optical complex-field letter recognition}\label{subsec2_3}

\begin{figure}[H]
\centering
\includegraphics[width=0.85\textwidth]{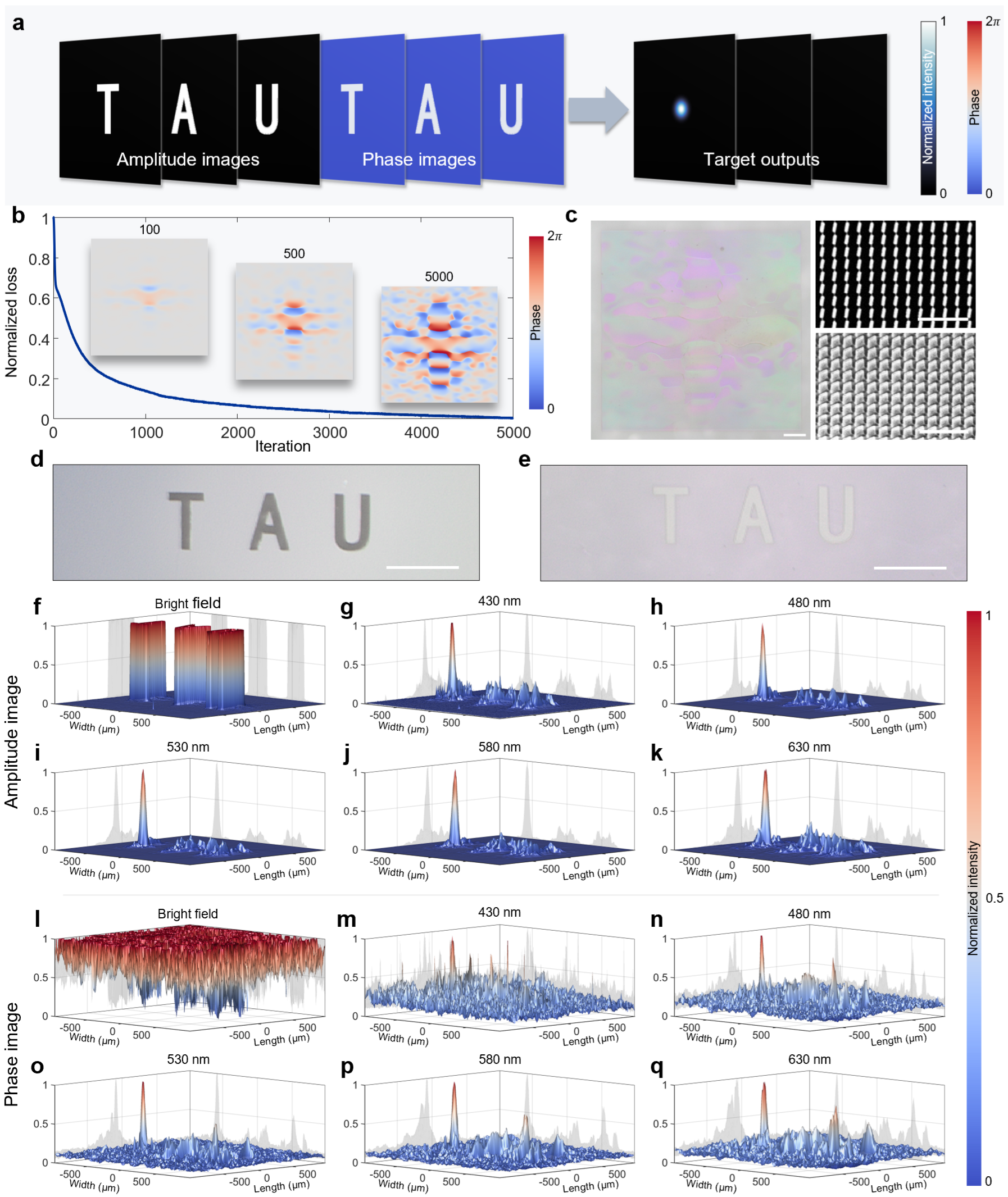}
\caption{\textbf{Experimental demonstration of meta-optical complex-field pattern recognition.} 
\textbf{a} Amplitude- and phase-encoded input images corresponding to the letters T, A, and U, along with the target output intensity distributions used for inverse design. The target pattern T is mapped to a centered Gaussian output intensity, while the distractor patterns (A and U) are assigned a uniform dark target output to suppress unwanted responses. 
\textbf{b} Optimization trajectory showing the normalized loss during 5000 iterations. Insets show the evolution of the metasurface phase profile from an initially flat distribution at iterations 100, 500, and 5000. 
\textbf{c} Optical microscope image of the fabricated metasurface and corresponding SEM images of the TiO$_2$ nanopillar array. 
\textbf{d}, \textbf{e} Amplitude and phase samples comprising the letters T, A, and U, used for pattern recognition testing (acquired in reflection mode). 
\textbf{f--k} Experimental recognition results for the amplitude sample, showing the unprocessed output (\textbf{f}) and the processed outputs at wavelengths of 430, 480, 530, 580, and 630~nm. 
\textbf{l--q} Corresponding recognition results for the phase sample, including the unprocessed output (\textbf{l}) and the processed outputs at the same wavelengths. 
Intensity distributions are projected onto the coordinate planes to compare output peak heights. Scale bars: 200~\textmu m (optical microscopy in \textbf{c}), 2~\textmu m (SEM in \textbf{c}), and 500~\textmu m (\textbf{d}, \textbf{e}).}
\label{fig 3}
\end{figure}

To further demonstrate the versatility of the proposed meta-optical framework, we next consider a letter recognition task. Compared with edge detection, this task places more demanding constraints on the processor, as it must selectively respond to a designated target pattern while suppressing signals from distractor inputs.

The metasurface is designed to recognize the letter T among the letters T, A, and U. The design objective is to generate a sharply localized intensity maximum when the input corresponds to the target letter T, while producing minimal output for the distractor letters A and U. Accordingly, the desired output for the target input is defined as a centered Gaussian intensity distribution, whereas the distractor inputs are assigned spatially uniform dark outputs. The input optical fields are constructed using the same amplitude- and phase-encoding schemes as in the edge-detection experiment, with the encoded input patterns and corresponding target intensity distributions shown in Figure~\ref{fig 3}a.

The metasurface phase profile is optimized by minimizing a composite loss function that enforces accurate formation of the target response while suppressing unwanted outputs associated with distractor inputs. Owing to the linear and energy-conserving properties of the optical system, the total output intensity is constrained by the input energy, and responses to distractors cannot be completely eliminated. Instead, the optimization redistributes residual energy as uniformly as possible to prevent the formation of localized intensity peaks. The target-shape loss is defined as
\begin{equation}
\mathcal{L}_{\mathrm{shape}} = \left\| I_{\mathrm{tar}}(x,y) - G(x,y) \right\|_2^2,
\end{equation}
where $I_{\mathrm{tar}}(x,y)$ denotes the output intensity distribution for the target input and $G(x,y)$ is a centered Gaussian function representing the desired response. To further suppress residual signals associated with distractor inputs, a cross-talk loss is introduced,
\begin{equation}
\mathcal{L}_{\mathrm{xtalk}} =
\frac{I_{\mathrm{dis}}^{\mathrm{peak}}}{I_{\mathrm{tar}}^{\mathrm{peak}}}
+ \alpha \left( \frac{I_{\mathrm{dis}}^{\mathrm{peak}}}{I_{\mathrm{dis}}^{\mathrm{mean}}} - 1 \right),
\end{equation}
where $I_{\mathrm{tar}}^{\mathrm{peak}}$ denotes the peak output intensity for the target input, $I_{\mathrm{dis}}^{\mathrm{peak}}$ is the average peak intensity across all distractor inputs, and $I_{\mathrm{dis}}^{\mathrm{mean}}$ is the corresponding average mean intensity. The first term penalizes residual peak responses from distractors relative to the target, while the second term discourages localized intensity concentrations by promoting spatially uniform background distributions, with the relative weighting controlled by $\alpha$. The total loss function is then given by
\begin{equation}
\mathcal{L}_{\mathrm{rec}} = \mathcal{L}_{\mathrm{shape}} + \beta \, \mathcal{L}_{\mathrm{xtalk}},
\end{equation}
where $\beta$ controls the trade-off between fidelity of the target response and suppression of unwanted outputs. Additional implementation details are provided in Supplementary Information Section~1.

The optimized letter-recognition processor is implemented as a square aperture comprising $4000 \times 4000$ meta-atoms, corresponding to a physical side length of 1.8\,mm. The optimization trajectory, shown in Figure~\ref{fig 3}b, shows a progressive decrease in the normalized loss over 5000 iterations, with representative phase profiles shown at intermediate and final stages. The fabricated metasurface is shown in Figure~\ref{fig 3}c.

The image samples used for experimental evaluation are shown in Figure~\ref{fig 3}d,e, corresponding to amplitude- and phase-encoded images comprising the letters T, A, and U (Supplementary Information Sections~3 and 13). Each sample includes both the target and distractor letters simultaneously, enabling direct assessment of the processor’s ability to selectively recognize the target pattern from complex-field inputs. Experimental results obtained using the setup shown in Figure~\ref{fig 1}e are presented in Figure~\ref{fig 3}f--q. For each encoding scheme, the unprocessed bright-field image is shown alongside the corresponding outputs after metasurface processing at wavelengths of 430, 480, 530, 580, and 630~nm. For the amplitude-encoded image, the unprocessed image clearly reveals all three letters. After processing, a pronounced intensity peak appears at the location of the target letter T, while the responses associated with the distractor letters A and U are strongly suppressed across the entire wavelength range. For the phase-encoded image, the unprocessed image exhibits minimal contrast under direct illumination. Following metasurface processing, a strong and spatially localized response emerges at the position of the target letter, whereas the distractor regions remain relatively uniform and dim. Although the background noise is slightly elevated at 430~nm owing to reduced spin-conversion efficiency and incomplete filtering of unmodulated light, the target signal remains clearly distinguishable throughout the measured spectral range.

\begin{figure}[H]
\centering
\includegraphics[width=0.85\textwidth]{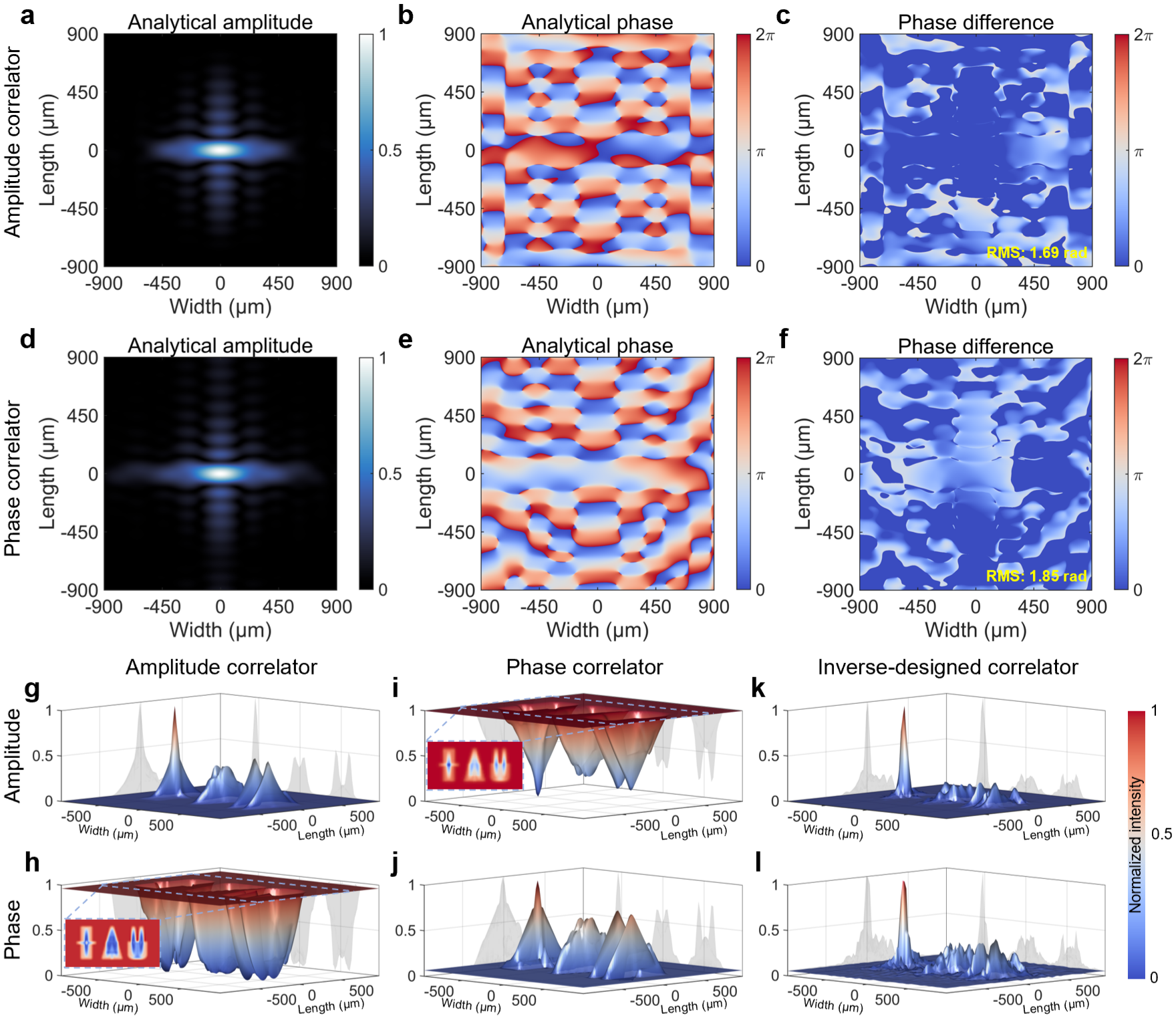}
\caption{\textbf{Comparison between analytical cross-correlation filtering and the inverse-designed meta-optical processor.}
\textbf{a} Amplitude of the complex transmission function of an analytical cross-correlation filter for an amplitude-encoded target letter T.
\textbf{b} Phase profile of the corresponding analytical filter.
\textbf{c} Phase difference between the inverse-designed metasurface and the analytical filter in \textbf{b}. A global phase offset is removed (RMS: 1.69\,rad).
\textbf{d} Amplitude of the complex transmission function of an analytical cross-correlation filter for a phase-encoded target letter T.
\textbf{e} Phase profile of the corresponding analytical filter.
\textbf{f} Phase difference between the inverse-designed metasurface and the analytical filter in \textbf{e} (RMS: 1.85\,rad).
\textbf{g}, \textbf{h} Simulated output intensities for amplitude- and phase-encoded TAU image inputs (Figure~\ref{fig 3}d,e), processed using the analytical cross-correlation filter designed for amplitude-encoded inputs. The inset in \textbf{h} shows a top-view projection of the intensity distribution.
\textbf{i}, \textbf{j} Corresponding simulated outputs obtained using the analytical filter designed for phase-encoded inputs. The inset in \textbf{j} shows a top-view projection of the intensity distribution.
\textbf{k}, \textbf{l} Simulated outputs obtained using the inverse-designed meta-optical processor in Figure~\ref{fig 3}b.}
\label{fig 4}
\end{figure}

This inverse-designed meta-optical processor exhibits improved selectivity and broad applicability to different input encodings compared with conventional optical pattern recognition approaches based on analytical cross-correlation filtering~\cite{yu2025meta,lugt1974coherent,goodman2017introduction}. For comparison, we construct analytical spatial filters that implement matched-filtering through cross-correlation with a predefined target pattern. Within this matched-filtering framework, the analytical filter is given by the complex-conjugated spectrum $\overline{\mathcal{F}\{T(x,y)\}}$, where $T(x,y)$ represents the complex optical field of the target, whether amplitude- or phase-encoded. This formulation yields a complex-valued transfer function that intrinsically requires simultaneous and accurate modulation of both amplitude and phase~\cite{yu2025meta}.

Figures~\ref{fig 4}a,b and~\ref{fig 4}d,e show the amplitude and phase profiles of analytical cross-correlation filters designed for amplitude- and phase-encoded target letters T, respectively. These are compared with the phase profile of the inverse-designed meta-optical processor shown in Figure~\ref{fig 3}b. The corresponding phase differences, after removal of a global phase offset, are presented in Figures~\ref{fig 4}c and~\ref{fig 4}f, yielding root-mean-square (RMS) differences of 1.69\,rad and 1.85\,rad for the amplitude- and phase-encoded targets. Although the inverse-designed phase profile partially resembles the analytical filters, it exhibits pronounced deviations at high spatial frequencies in both cases, as well as in the low-frequency components of the phase-encoded filter. These deviations are critical for suppressing background intensity and enabling robust discrimination across different input modalities, which cannot be achieved by either analytical filter.

The resulting recognition performance is illustrated by the simulated results in Figures~\ref{fig 4}g--l. When the analytical filter designed for an amplitude-encoded target is applied to an amplitude input (Figure~\ref{fig 4}g), the target letter T is detected; however, the distractor letters A and U also generate strong responses, indicating poor selectivity. Applying the same filter to a phase-encoded input yields no discernible target response and a high background intensity (Figure~\ref{fig 4}h). Conversely, the analytical filter designed for a phase-encoded target fails to produce a meaningful response for amplitude inputs (Figure~\ref{fig 4}i) and provides only weak enhancement at the target location for phase inputs, with limited suppression of distractors (Figure~\ref{fig 4}j). In contrast, the inverse-designed meta-optical processor consistently produces a strong and spatially localized response at the target position for both amplitude- and phase-encoded inputs (Figures~\ref{fig 4}k,l), while effectively suppressing distractor signals, consistent with the experimental results shown in Figures~\ref{fig 3}f--q. Moreover, unlike analytical cross-correlation filters, which fundamentally rely on complex-valued modulation, the inverse-designed meta-optical processor achieves high selectivity and robustness using a phase-only metasurface. This simplifies physical implementation while enabling efficient and broadband operation~\cite{arbabi2015dielectric,overvig2019dielectric,ren2020complex,yu2025meta}, highlighting how inverse design overcomes the constraints of analytically prescribed optical filters to enable practical, encoding-robust image processing.

\subsection*{Compact meta-optical processor with integrated imaging and computation}\label{subsec2_4}

\begin{figure}[H]
\centering
\includegraphics[width=0.85\textwidth]{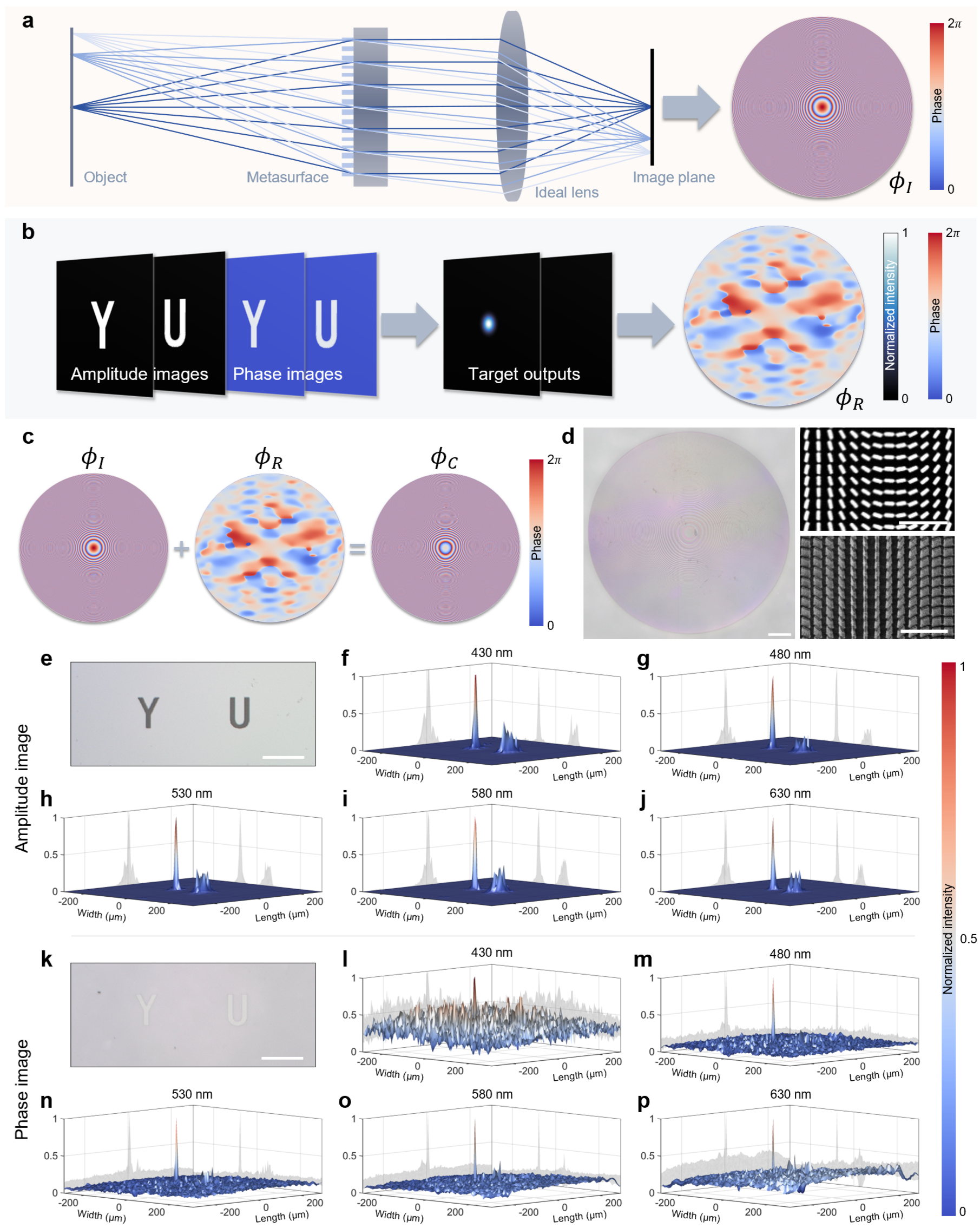}
\caption{\textbf{Compact meta-optical processor integrating imaging and pattern recognition.}
\textbf{a} Imaging phase $\phi_I$ designed using ray tracing for the compact meta-optical processor configuration.
\textbf{b} Amplitude- and phase-encoded input images corresponding to the target letter Y and the distractor letter U, together with the associated target output intensity distributions and the optimized recognition phase profile $\phi_R$.
\textbf{c} Composite phase profile $\phi_C$, obtained by combining the imaging phase $\phi_I$ with the pattern recognition phase $\phi_R$.
\textbf{d} Optical microscope image of the fabricated metasurface and corresponding SEM images of the TiO$_2$ nanopillar array.
\textbf{e}, \textbf{k} Amplitude and phase samples comprising the letters Y and U.
\textbf{f--j}, \textbf{l--p} Corresponding processed outputs at wavelengths of 430, 480, 530, 580, and 630~nm for the amplitude- and phase-encoded inputs, respectively.
Scale bars: 200\,\textmu m (optical microscopy in \textbf{d}), 2\,\textmu m (SEM in \textbf{d}), 500\,\textmu m (\textbf{e}, \textbf{k}).}
\label{fig 5}
\end{figure}

To move the meta-optical processor toward practical deployment, we develop a compact architecture that integrates optical computation and image formation within a single metasurface layer. Although the demonstrations presented here focus on pattern recognition, the same strategy can be readily extended to other analog optical image processing tasks. In this configuration, the meta-optical processor is placed at the spatial-frequency plane of the input image, where the metasurface encodes a composite phase profile that combines an imaging phase with a recognition phase. This approach enables direct transformation from the input image spatial spectrum to a processed output intensity, eliminating the need for a conventional multi-element $4f$ optical system~\cite{goodman2017introduction}. Although a single lens is retained to relay the input image field onto the metasurface aperture, the overall architecture substantially reduces system complexity and footprint, enabling a compact and practical implementation. Additional details are provided in Supplementary Information Section~10.

The compact meta-optical processor is implemented as a circular aperture with a diameter of 1.8\,mm. To achieve high-quality image formation over a wide field of view, the imaging phase profile $\phi_I$ is designed using reverse ray-tracing simulations, as illustrated in Figure~\ref{fig 5}a~\cite{malacara2003handbook}. In this approach, the desired image plane is modeled as an object composed of point sources spanning a field height of up to 1\,mm, and rays are traced backward to the metasurface located 9\,mm away, corresponding to the designed imaging focal length. An ideal thin lens placed after the metasurface refocuses the collimated rays onto the output image plane, enabling aberration-free evaluation of the imaging performance. The imaging phase profile $\phi_I$ is expressed as an even-order radial polynomial in the normalized aperture radius $r$ and is optimized by minimizing the RMS spot size across the field,
\begin{equation}
\phi(r) = -526.31\,r^2 - 9.14\,r^4 + 13.23\,r^6 - 1.15\,r^8 - 0.93\,r^{10} - 2.58\,r^{12}.
\label{eq:radial_phase}
\end{equation}
Further details of the imaging phase optimization are provided in Supplementary Information Section~6. For demonstration, the recognition phase $\phi_R$ is designed using the inverse-design framework described earlier to distinguish the target letter Y from the distractor letter U. Figure~\ref{fig 5}b shows the amplitude- and phase-encoded inputs, the corresponding target outputs, and the resulting optimized phase profile (see Supplementary Information Section~7 for additional details). The final metasurface implements a composite phase profile $\phi_C = \phi_I + \phi_R$, as illustrated in Figure~\ref{fig 5}c. The fabricated metasurface corresponding to the composite phase design is shown in Figure~\ref{fig 5}d.

The performance of the compact meta-optical processor is assessed experimentally using amplitude and phase samples comprising the letters Y and U (Figure~\ref{fig 5}e,k; see Supplementary Information Sections~3 and 13). Measurements are performed using the setup shown in Figure~\ref{fig 1}f (see Methods and Supplementary Information Section~11). Figures~\ref{fig 5}f--j and~\ref{fig 5}l--p present the corresponding processed output intensities at wavelengths of 430, 480, 530, 580, and 630~nm for amplitude- and phase-encoded inputs, respectively. In both cases, the metasurface produces a strong and spatially localized response at the position of the target letter Y, while responses associated with the distractor letter U are effectively suppressed across the entire spectral range. These results confirm that the compact architecture preserves the recognition performance achieved with the larger $4f$ optical system, despite the substantially reduced optical footprint. For phase-encoded inputs, a modest increase in background intensity is observed at 430~nm, which can be attributed to reduced spin-conversion efficiency and residual unmodulated light at shorter wavelengths; nevertheless, the target response remains clearly distinguishable throughout the measured bandwidth. Additional simulated and experimental results are provided in Supplementary Information Sections~8 and 9.

\section*{Discussion}\label{sec3}

In summary, this work demonstrates a broadband meta-optical processor that performs analog image processing directly on complex optical fields, supporting both amplitude- and phase-encoded inputs within a unified metasurface framework. The processor is implemented using a dielectric metasurface whose phase profile is optimized through an end-to-end, task-driven inverse design approach, enabling wavefront transformations tailored to specific image processing tasks. Using this platform, we experimentally realize edge detection and pattern recognition across a 430--630\,nm wavelength range, demonstrating broadband operation in the visible spectrum.

Beyond demonstrating individual image processing functions, this work highlights the flexibility of inverse-designed metasurfaces as a general platform for analog optical computation. In contrast to many previously reported metasurface-based processors that are limited to a specific task, a narrow operating bandwidth, or a single input modality, the proposed approach supports diverse processing functions and operates consistently for both amplitude- and phase-encoded images across a broad spectral bandwidth. Table~\ref{tab:comparison} summarizes representative prior works and places the present platform in context, highlighting complementary advances in functionality, spectral bandwidth, and input versatility.

\begin{table}[h]
\caption{Comparison of representative metasurface-based analog image processing approaches.}

\label{tab:comparison}
\begin{tabular*}{\textwidth}{@{\extracolsep\fill}l l c c c l}
\toprule
\textbf{Ref.} & \textbf{Functionality} & \textbf{4$f$} & \textbf{Image} & \textbf{Spec.} & \textbf{Principle} \\
\midrule


Ref.~\cite{zhou2019optical,zhou2021two,yu2025multifunctional} & \multirow{4}{*}{Edge detection} & $\checkmark$ & A/P & B & \multirow{2}{*}{Diff. interference} \\
\cmidrule(lr){1-1} \cmidrule(lr){3-5}
Ref.~\cite{wang2023single} &  & $\times$ & A/P & S \\
\cmidrule(lr){1-1} \cmidrule(lr){3-6}
Ref.~\cite{huo2020photonic,kim2022spiral,huo2024broadband} & & $\checkmark$/$\times$ & A/P & B & Phase contrast \\
\cmidrule(lr){1-1} \cmidrule(lr){3-6}
Ref.~\cite{kwon2018nonlocal,guo2018photonic,zhou2020flat} & & $\times$ & A/P & S & \multirow{2}{*}{Nonlocal resonance} \\
\cmidrule(lr){1-5}
Ref.~\cite{fu2022ultracompact} & High/Low-pass filt. & $\times$ & A/P & M & \\
\cmidrule(lr){1-6}

Ref.~\cite{wang2022single,wang2023single,yu2025meta} & Multi-task & $\checkmark$/$\times$ & A & S & Cplx.-amp. filt. \\
\cmidrule(lr){1-6}

\ding{72} This work & Multi-task & $\checkmark$/$\times$ & A/P & B & Phase-only filt. \\
\botrule
\end{tabular*}

\vspace{1ex}
\footnotesize{
\textbf{4$f$}: use of a conventional $4f$ optical system; 
\textbf{Image}: supported input encoding; 
\textbf{A}: amplitude-encoded; 
\textbf{P}: phase-encoded; 
\textbf{Spec.}: spectral operating range; 
\textbf{S}: single wavelength; 
\textbf{M}: multiple discrete wavelengths; 
\textbf{B}: broadband.
}
\end{table}

We further introduce a compact processor architecture that integrates optical computation and image formation within a single metasurface layer, eliminating the need for conventional multi-element $4f$ configurations and substantially reducing the system footprint. Experimental results show that this compact implementation preserves the essential processing performance achieved in larger optical systems, while providing a more practical route toward deployment in space- and complexity-constrained optical environments.

Overall, this work establishes inverse-designed meta-optical processors as a compact and flexible approach for analog image processing on complex optical fields, with clear potential for integration in practical imaging and optical computing systems~\cite{hu2024diffractive,hu2024metasurface}.

\section*{Methods}\label{sec4}

\subsection*{Simulation methods}\label{subsec4_1}
The complex transmission responses of meta-atoms were simulated using the RF Module of COMSOL Multiphysics (v6.2). Physical modeling and phase-profile optimization of the meta-optical processor were based on Fourier optics~\cite{goodman2017introduction} and implemented in Python (v3.10) using PyTorch (v2.0.0) with CUDA (v11.7), enabling automatic differentiation and gradient-based optimization. Ray-tracing simulations for the design of the imaging phase profile were performed using Ansys Zemax OpticStudio (2024 R1.00). Experimental data analysis and additional numerical simulations presented in Figure~\ref{fig 5} and the Supplementary Information were carried out in MATLAB (R2022b). Further methodological details are provided in Supplementary Information Sections~1, 2, and 8.

\subsection*{Metasurface fabrication}\label{subsec4_2}
The metasurfaces were fabricated from a 600\,nm-thick TiO$_2$ film deposited on a fused silica substrate by ion-beam sputtering. A polymethyl methacrylate electron-beam resist and a Al charge-dissipation layer were sequentially applied. Following electron-beam lithography exposure and development, a chromium (Cr) hard mask was defined by metal evaporation and lift-off. The nanopillar patterns were then transferred into the TiO$_2$ layer using reactive-ion etching. After pattern transfer, the Cr hard mask was removed by wet chemical etching, yielding the final nanopillar metasurfaces. Detailed fabrication procedures are provided in Supplementary Information Section~12.

\subsection*{Image samples fabrication}\label{subsec4_3}
Amplitude- and phase-encoded image samples were fabricated on fused silica substrates using laser direct writing lithography. Phase samples were realized by patterning a negative photoresist layer, followed by development, to introduce spatially varying optical phase delays through thickness modulation. Amplitude masks were fabricated by resist patterning followed by deposition of a Cr film and lift-off, yielding spatially varying transmission features. Additional fabrication details are provided in Supplementary Information Section~13.

\subsection*{Optical characterization setups}\label{subsec4_4}
Experimental characterization of the meta-optical processors was performed using two optical setups, schematically shown in Figure~\ref{fig 1}e,f. For edge detection and letter T recognition, the setup in Figure~\ref{fig 1}e was used. A collimated laser beam illuminated the image samples. A circular polarizer, consisting of a linear polarizer followed by a quarter-wave plate, converted the incident beam into a left-handed circularly polarized (LCP) state. Lenses~1 and~2 formed a $4f$ imaging system, with the meta-optical processor placed at the Fourier plane to modulate the spatial-frequency components of the input image. A matched circular polarizer, configured as an analyzer, was positioned after the metasurface to selectively transmit the cross-polarized right-handed circularly polarized (RCP) output. The processed image was recorded using a camera.

For the compact meta-optical processor used in the letter Y recognition task, the optical setup shown in Figure~\ref{fig 1}f was used. In this configuration, the metasurface simultaneously performed spatial-frequency modulation and image formation. The input image was placed at the front focal plane of a single lens, while the metasurface was positioned at its back focal plane to generate the processed output on the image sensor. Additional experimental details are provided in Supplementary Information Section~11.

\backmatter

\bmhead{Data availability}
The data supporting this study are not publicly available but can be provided by the corresponding author upon reasonable request.

\bmhead{Supplementary information}
Supplementary information is available.

\bmhead{Acknowledgements}
L.Y. and H.C. acknowledge financial support from the European Union’s Horizon 2020 research and innovation programme under the Marie Skłodowska-Curie grant agreement No 956770.

\bibliography{sn-bibliography}

\end{document}